\documentclass[]{spie}  %>>> use for US letter paper
\usepackage[]{graphicx}

\title{Design and characterization of 90 GHz feedhorn-coupled TES polarimeter pixels in the SPTpol camera} 

\author{
J.T.~Sayre\supit{a}, P.~Ade\supit{b}, K.A.~Aird\supit{r}, J.E.~Austermann\supit{c}, J.A.~Beall\supit{d}, D.~Becker\supit{d}, B.A.~Benson\supit{a,s}, L.E.~Bleem\supit{a,q}, J.~Britton\supit{d}, J.E.~Carlstrom\supit{a,e,q,s,t}, C.L.~Chang\supit{a,e,s}, H-M.~Cho\supit{d}, T.M.~Crawford\supit{a,t}, A.T.~Crites\supit{a,t}, A.~Datesman\supit{g}, T.~de Haan\supit{f}, M.A.~Dobbs\supit{f}, W.~Everett\supit{m}, A.~Ewall-Wice\supit{a,q}, E.M.~George\supit{h}, N.W.~Halverson\supit{c,p}, N.~Harrington\supit{h}, J.W. ~Henning\supit{c}, G.C.~Hilton\supit{d}, W.L.~Holzapfel\supit{h}, J.~Hubmayr\supit{d}, K.D.~Irwin\supit{d}, M.~Karfunkle\supit{a,q}, R.~Keisler\supit{a,q,s}, J.~Kennedy\supit{f}, A.T.~Lee\supit{h}, E.~Leitch\supit{m}, D.~Li\supit{d}, M.~Lueker\supit{j}, D.P.~Marrone\supit{o}, J.J.~McMahon\supit{k}, J.~Mehl\supit{a,s}, S.S.~Meyer\supit{a,q,s,t}, J.~Montgomery\supit{a,q}, T.E.~Montroy\supit{a}, J.~Nagy\supit{a}, T.~Natoli\supit{a,q}, J.P.~Nibarger\supit{d}, M.D.~Niemack\supit{d}, V.~Novosad\supit{g}, S.~Padin\supit{m}, C.~Pryke\supit{l}, C.L.~Reichardt\supit{h}, J.E.~Ruhl\supit{a}, B.R.~Saliwanchik\supit{a}, K.K.~Schaffer\supit{n}, E.~Shirokoff\supit{j}, K.~Story\supit{a,q}, C.~Tucker\supit{b}, K.~Vanderlinde\supit{f}, J.D.~Vieira\supit{j}, G.~Wang\supit{e}, R.~Williamson\supit{a,s}, V.~Yefremenko\supit{e,g}, K.~W.~Yoon\supit{d}, E.~Young\supit{h}
\skiplinehalf
\supit{a} Case Western Reserve University, Cleveland, OH 44106, USA
\skiplinehalf
\supit{b} Cardiff School of Physics and Astronomy, Cardiff University, Cardiff, United Kingdom
\skiplinehalf
\supit{c} Department of Astrophysical and Planetary Sciences, University of Colorado, Boulder, CO 80309, USA
\skiplinehalf
\supit{d} NIST, Boulder, CO 80305, USA
\skiplinehalf
\supit{e} High Energy Physics Division, Argonne National Laboratory, Argonne, IL 60439, USA
\skiplinehalf
\supit{f} McGill University, Montreal, Quebec, Canada
\skiplinehalf
\supit{g} Materials Science Division, Argonne National Laboratory, Argonne, IL 60439, USA
\skiplinehalf
\supit{h} University of California, Berkeley, 151 LeConte Hall Berkeley, CA 94720, USA
\skiplinehalf
\supit{j} California Institute of Technology, Pasadena, CA 91125, USA
\skiplinehalf
\supit{k} University of Michigan, Ann Arbor, MI, USA
\skiplinehalf
\supit{l} University of Minnesota, Minneapolis, MN 55455, USA
\skiplinehalf
\supit{m} Kavli Institute for Cosmological Physics, Department of Physics, Enrico Fermi Institute, The University of Chicago, Chicago, IL 60637, USA
\skiplinehalf
\supit{n} School of the Art Institute of Chicago, Chicago, IL 60603, USA
\skiplinehalf
\supit{o} Steward Observatory, University of Arizona, 933 North Cherry Avenue, Tucson, AZ 85721, USA
\skiplinehalf
\supit{p} Department of Physics, University of Colorado, Boulder, CO 80309, USA
\skiplinehalf
\supit{q} Department of Physics, University of Chicago, 5640 South Ellis Avenue, Chicago, IL 60637, USA
\skiplinehalf
\supit{r} University of Chicago, 5640 South Ellis Avenue, Chicago, IL 60637, USA
\skiplinehalf
\supit{s} Enrico Fermi Institute, University of Chicago, 5640 South Ellis Avenue, Chicago, IL 60637, USA
\skiplinehalf
\supit{t} Department of Astronomy and Astrophysics, University of Chicago, 5640 South Ellis Avenue, Chicago, IL 60637, USA
}

\begin{document} 
\maketitle 

%%%%%%%%%%%%%%%%%%%%%%%%%%%%%%%%%%%%%%%%%%%%%%%%%%%%%%%%%%%%% 
\begin{abstract}
The SPTpol camera is a two-color, polarization-sensitive bolometer receiver, and was installed on the 10 meter South Pole Telescope in January 2012. SPTpol is designed to study the faint polarization signals in the Cosmic Microwave Background, with two primary scientific goals. One is to constrain the tensor-to-scalar ratio of perturbations in the primordial plasma, and thus constrain the space of permissible inflationary models. The other is to measure the weak lensing effect of large-scale structure on CMB polarization, which can be used to constrain the sum of neutrino masses as well as other growth-related parameters. The SPTpol focal plane consists of 
seven 84-element monolithic arrays of 150~GHz pixels (588 total) and 180 individual 90 GHz single-pixel modules.  In this paper we present the design and characterization of the 90~GHz modules.
\end{abstract}


\section{INTRODUCTION}
\label{sec:intro}  
The current state of Cosmic Microwave Background radiation (CMB) research is in transition from primarily studying temperature fluctuations (of order $100 \mu$K rms) to also mapping the faint CMB polarization signals (of order few $\mu$K rms) on the sky.  Two major cosmological results are expected to come from analyzing maps of CMB polarization.  The first is either a detection or an upper limit on the amplitude of primordial gravity waves, which would provide a direct probe of the era of inflation and offer the first concrete insight into the energy scale at which inflation occurred.  The second is a measurement of the weak lensing signal of CMB radiation due to the gravitational effect of large scale structure between us and the surface of last scattering.  This weak lensing signal provides a means of constraining neutrino mass through its effect on the history of 
structure growth.

The South Pole Telescope underwent an upgrade in the austral summer of 2011-2012, during which a new, polarization-sensitive camera was installed.  The upgraded telescope and new camera, collectively referred to as SPTpol, is among the most sensitive CMB polarimeters currently in operation.  The SPTpol focal plane consists of seven 84-pixel monolithic polarization sensitive feedhorn-coupleed detector arrays at 150~GHz (588 pixels, 1176 detectors total), and 
180 individual feedhorn-detector modules (180 pixels, 360 detectors total) at 90~GHz.  Each 90 GHz pixel is a self-contained module consisting of a feedhorn and waveguide piece mounted to a base containing two independent, crossed bolometers with connections for readout and physical mounting to the focal plane.  The bolometers couple via PdAu resistive resonators to a single circular waveguide mode.  The resistive resonators are mounted on a SiN island along with MoAu transition edge sensors (TES) used for readout.  In this paper, we discuss the thermal, electrical, and optical design of the 
90~GHz pixels, and review the pre-deployment characterization of their properties.

\section{DESIGN OF 90 GHZ PIXELS} 
\label{sec:Design}

\subsection{Optical Design}
Polarization transmission fidelity, azimuthal beam symmetry, and maximum optical coupling efficiency are the primary optical design considerations for the 90 GHz detector modules.  We use a smoothwall profiled-feed to maximize beam symmetry for a linearly polarized detector, in a simple and easy-to-machine design compared to corrugated feeds \cite{Zeng}.  Symmetric beams are required for good common-mode subtraction between polarized detectors oriented at $90^{\circ}$ to each other, which is essential for minimizing leakage of temperature anisotropies into differenced-detector polarization signals.  Metal mesh low-pass filters in the optical system skyward of the feedhorns define the upper band edge and block high frequencies, while the lower edge of the band is defined by the 2.35-mm waveguide diameter at the back end of the feed.  Radiation is coupled from the waveguide to the TES island through lossy PdAu bar resonators deposited on a SiN membrane.  

Initial designs for the optical elements of the 90 GHz detectors were generated using HFSS simulations to model the optical performance of a mounted single-bar absorber \cite{McMahon}.  We chose a long, thin bar as the absorber shape to optimize coupling to the TE11 spatial mode supported in the waveguide.  Initial prototypes had absorbers made of Au/Cr bilayers, but the material was changed to PdAu to allow for a good impedance match between the absorber and waveguide fields with a thinner and thus easier-to-manufacture layer.  Simulations informed the absorber geometry (as summarized in 
Table~ \ref{tab:geoms}) and location within the waveguide that maximized coupling between incident radiation and the absorber at the center of the optical band.   The simulations predict co-polar coupling of 93\% averaged across the band, a cross-polar coupling of less than 0.1\%, and confirmed that the second (TM01) mode that turns on near the upper edge of the band does not couple appreciably to the absorber structure \cite{McMahon}.  Further HFSS simulations were used to establish acceptable limits of detector misalignment, and showed that minimal performance loss resulted from radial translations within the waveguide up to 75 $\mu$m, rotational misalignment of less than $2^{\circ}$, and changes in distance between the aborbers and backshort of less than 100 $\mu$m.  In the final configuration, the two single-detector wafers are mounted face-to-face at a separation of ~25 $\mu$m, achieved by laying 25 $\mu$m diameter wirebonds between the detectors' Si support structures.

   \begin{figure}
   \begin{center}
   \begin{tabular}{c}
   \includegraphics[height=4cm]{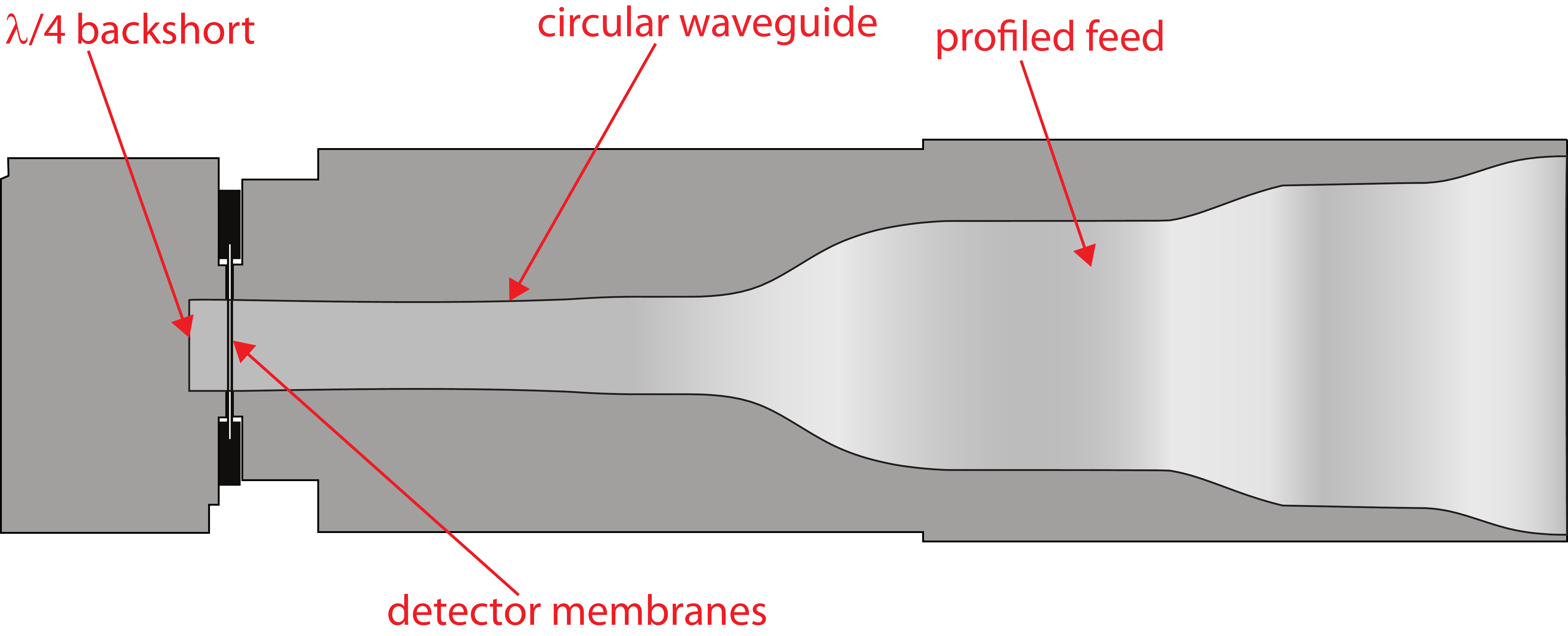}
   \end{tabular}
   \end{center}
   \caption[Full pixel] 
   { \label{fig:pixel} 
Design of a 90 GHz pixel module.  Low-pass filters sit in front of the horn aperture to define the upper edge of the band.  The profiled feed provides excellent beam symmetry without the complication and expense inherent in manufacturing traditional corrugated feeds.  The circular waveguide TE11 mode cutoff defines the lower edge of the band, and the detectors sit a quarter wavelength (at the band center) in front of the terminating backshort.  Wiring comes out of the module radially at the position of the Si detector frames.
}   \end{figure}

\begin{table}[h]
\caption{Nominal dimensions of 90 GHz optical structures from HFSS simulations} 
\label{tab:geoms}
\begin{center}       
\begin{tabular}{|l|r|} %% this creates two columns
\hline
\rule[-1ex]{0pt}{3.5ex}  Absorber width & 18 $\mu$m  \\
\hline
\rule[-1ex]{0pt}{3.5ex}  Absorber length & 1300 $\mu$m  \\
\hline
\rule[-1ex]{0pt}{3.5ex}  Distance between absorbers & 25 $\mu$m  \\
\hline
\rule[-1ex]{0pt}{3.5ex}  Waveguide diameter & 2350 $\mu$m  \\
\hline
\rule[-1ex]{0pt}{3.5ex}  Distance from absorber to backshort & 1080 $\mu$m  \\
\hline
\rule[-1ex]{0pt}{3.5ex}  Width of waveguide break & 150 $\mu$m  \\
\hline
\rule[-1ex]{0pt}{3.5ex}  Width of choke  & 840 $\mu$m \\
\hline
\rule[-1ex]{0pt}{3.5ex}  Distance from TES to waveguide & 150 $\mu$m \\
\hline 
\end{tabular}
\end{center}
\end{table} 

\subsection{Thermal  design}
Our detectors see a baseline optical loading background of photons from the CMB, atmosphere, and various emissive instrument elements in the optical path.  They are also biased with electrical power via the resistive TES.  For a thermal link with a power law temperature dependence, the total quiescent power deposited on the TES thermal island from both optical and electrical bais raises its temperature above the cryogenic bath according to
\begin{equation}\label{eqn:heattrans} 
P_{total}=\kappa (T_{TES}^n - T_{bath}^n),
\end{equation}
where $P_{total}$ is known as the saturation power when it is sufficient to drive $T_{TES}$ into its transition, and $\kappa$ depends on the materials and geometry of the TES and thermal island.  The background optical loading from CMB, atmosphere, and instrumental emission for SPTpol is $\sim10$pW, which represents the theoretical minimum saturation power for a TES detector.  Electrical power comes from the dissipation of readout current in the resistive TES and is easily tunable to provide a desired saturation power based on the TES design.  For fixed optical power loading, increased electrical bias power allows for more stable, linear operation over a wider range of observing conditions by reducing the impact of optical power fluctuations on the TES response, but also leads to increased noise due to a larger $\kappa$.  In SPTpol, the target electrical-to-optical power ratio at the TES transition temperature, $T_C$, is ~2.5:1, which represents a safety margin to ensure stable operation.  Saturation powers of ~25 pW at the 300 mK bath temperature set by SPTpol's He10 cryogenic cooler, result in a target $T_C$ of 510 mK.  Design dimensions for these thermal performance targets are listed in Table \ref{tab:therm_geoms}.  

Normal resistance of the TES depends on the thickness of the Au layer, and is constrained by readout requirements in SPTpol to be $\sim 1 \ \Omega$, while $T_C$ is determined by the ratio of the Mo and Au layer thicknesses\cite{Yefremenko}.  The target $T_C$ of 510 mK and target $R_n$ of 1 $\Omega$ result in typical thicknesses of (20 nm)/(30 nm) for the Mo/Au layers.  Both the TES and the PdAu bar absorber are deposited on a 1 $\mu$m thick film of SiN, which is patterned into a rectangular thermal mass with 8 long, thin legs connecting it to a support frame at the thermal bath temperature.  Thermal conductivity between the island and the bath is set by the geometry of the legs, which are each 10 $\mu$m wide, 1 $\mu$m thick, and 1350 $\mu$m long for a target saturation power of 25 pW .  Electrical leads to the TES in the form of Nb traces run along two of the legs to the TES but do not contribute appreciably to the thermal connection\cite{Wang}.  

\begin{table}[h]
\caption{Dimensions of 90 GHz thermal structures} 
\label{tab:therm_geoms}
\begin{center}       
\begin{tabular}{|l|r|} %% this creates two columns
\hline
\rule[-1ex]{0pt}{3.5ex}  SiN film thickness & 1 $\mu$m  \\
\hline
\rule[-1ex]{0pt}{3.5ex}  thermal island length x width & 3300 $\mu$m x 200 $\mu$m\\
\hline
\rule[-1ex]{0pt}{3.5ex}  thermal standoff leg length x width & 1350 $\mu$m  x 10 $\mu$m  \\
\hline
\rule[-1ex]{0pt}{3.5ex}  TES length x width & 70$\mu$m x 40 $\mu$m \\
\hline
\end{tabular}
\end{center}
\end{table} 
Steady-state performance as discussed above determines the TES bias parameters, but response to small, time-dependent signal fluctuations resulting from the telescope scanning across a CMB field is also a key design criterion for these detectors.  
At a bias power, $P_0$, and resulting TES temperature and resistance, $T_0$ and $R_0$, this response is parameterized by the dynamic thermal conductivity,  
\begin{equation}\label{eqn:G_dyn}
G_{dyn} = \frac{\partial P_{0}}{\partial T_{TES}}, 
\end{equation}
between the TES and the bath, and electro-thermal loopgain, $\mathcal{L}$, of response to optical power fluctuations defined as
\begin{equation}\label{eqn:loopgain}
\mathcal{L} = \frac{P_0 \alpha}{G_{dyn} T_0},
\end{equation} 
where
\begin{equation}\label{eqn:alpha}
\alpha = \left(\frac{T_0}{R_0}\right) \left(\frac{\partial \log R}{ \partial \log T} \right).
\end{equation}

Response to a time-varying signal depends on the optical time constant, $\tau_{opt}$, which reflects the time scale of thermalization between the absorber and the SiN island, and the TES thermal time constant, $\tau_0 = C_{TES}/G_{dyn}$, where $C_{TES}$ is the TES heat capacity.  SPTpol 90 GHz detectors have $\tau_{opt} << \tau_0$, such that overall detector reponse is dominated by the latter.  

We operate our detectors with a constant voltage bias applied across each TES, which results in their operating in negative electrothermal feedback (ETF) mode.  With negative ETF, an optical power fluctuation in the absorber changes the TES temperature and thus its resistance, inducing an electrical power fluctuation in the opposite direction of the optical power fluctuation and driving the TES back towards its bias point\cite{Irwin}.  The strength of this compensatory effect is parameterized by the ETF loopgain from (\ref{eqn:loopgain}), which is tuned to meet requirements for detector sensitivity, operational stability, and linearity.  The loopgain speeds up the thermal response time of the TES to
\begin{equation} \label{eqn:LTau}
\tau_{eff}=\frac{\tau_0}{(\mathcal{L}+1)},
\end{equation}
where $\tau_{eff}$ is the effective time constant of the feedback power fluctuation.  Higher loopgain and a reduced time constant im prove linearity and sensitivity are improved over a wider range of fluctuation amplitudes and frequencies.  However, stability depends on the bandwidth of loopgain-induced compensatory fluctations being smaller than the effective bandwidth of the detector readout scheme, as discussed in section \ref{sec:elec_design}.

\subsection{Electrical Design}\label{sec:elec_design}

SPTpol detectors are read out with a digital frequency-multiplexed (DfMux) readout system, in which each TES is biased at a frequency between 400 kHz and 1.2 MHz, with a comb of 12 multiplexed detectors sharing electrical bias lines\cite{Dobbs}.  CMB signals appear as small modulations between roughly 0.1 - 20 Hz in the detector time-ordered data, depending on the telescope scanning speed and on-sky scale of the CMB fluctuations.  Each TES is wired in series with a 24 $\mu$H inductor and a capacitor, C, that sets its bias frequency at $\omega = 1/\sqrt{LC}$, where the impedance of the L and C components of the circuit cancel each other, placing the full bias amplitude across the TES resistance.  A comb of 12 parallel LCR TES channels is placed in parallel with a 30 m$\Omega$ shunt resistor, which sets the voltage across the entire comb.   Current through the detector comb is recombined and read out by a SQUID amplifier, before being demodulated by warm electronics\cite{Dobbs}.  Resonant peaks are nominally spaced 60 kHz apart.   

In the DfMux readout system, the frequency dependent impedance of the resonant circuits used in the bias comb places limits on the useable signal bandwidth for each detector.  At the bias peak, where the signal frequency is $\omega$, the impedance of the series LC portion of circuit is zero, and current through the TES is simply $V_{bias}/R_{TES}$.  Modulations in TES resistance result in amplitude modulations of the carrier current that show up as sidebands to the carrier, and are thus attenuated by the nonzero off-resonance impedance of the bias circuit.  Impedance for each leg of the network is
\begin{equation} \label{eqn:Ztes}
Z_{leg} = R_{TES}+i \omega L - \frac{i}{\omega C}.
\end{equation} 
Signals that modulate the carrier amplitude at a frequency $\delta \omega$ are attenuated by the dependence of $Z_{leg}$ on $\omega$, which serves to reduce the amplitude and shift the phase of these off-resonance signals.  The signal bandwidth around each resonant peak is parameterized by the L/R time constant, $\tau_{elec}$, which is 28$\mu$s with 24 $\mu$H inductors and biased $R_{TES}$ of ~0.8 $\Omega$.  In addition to reducing the magnitude of carrier amplitude modulations, the LC circuit phase-shifts them, which results in TES instability if the electrothermal feedback has not been suppressed before significant interaction with the electrical filtering impedances.  The stability criterion relating the electrical and thermal time constants is $\tau_{eff} \geq 5.8 \tau_{elec}$ \cite{Lueker, Irwin2}.  Thus, for our readout system, a critical design criterion is that $\tau_{eff} > 0.17$ ms, which sets an upper limit on $\mathcal{L}$.

Initial prototypes for the SPTpol detectors consisted of a bare TES as described above, with geometry matching target steady-state thermal parameters.  However, these designs resulted in values of $\alpha$ and $\tau_0$ that were insufficient to maintain stability and linearity under expected operating conditions.  Using a solution implemented in the previous phase of SPT operations (SPT-SZ), we added a PdAu layer (called bling) that increases the heat capacity of the TES.  After several prototype designs, we settled on an L-shaped bling made of 400 nm thick PdAu with a volume of ~5000 $\mu$m$^3$ for a target $\tau_0$ of ~10ms.  

\begin{figure}
   \begin{center}
   \begin{tabular}{c}
   \includegraphics[height=7cm]{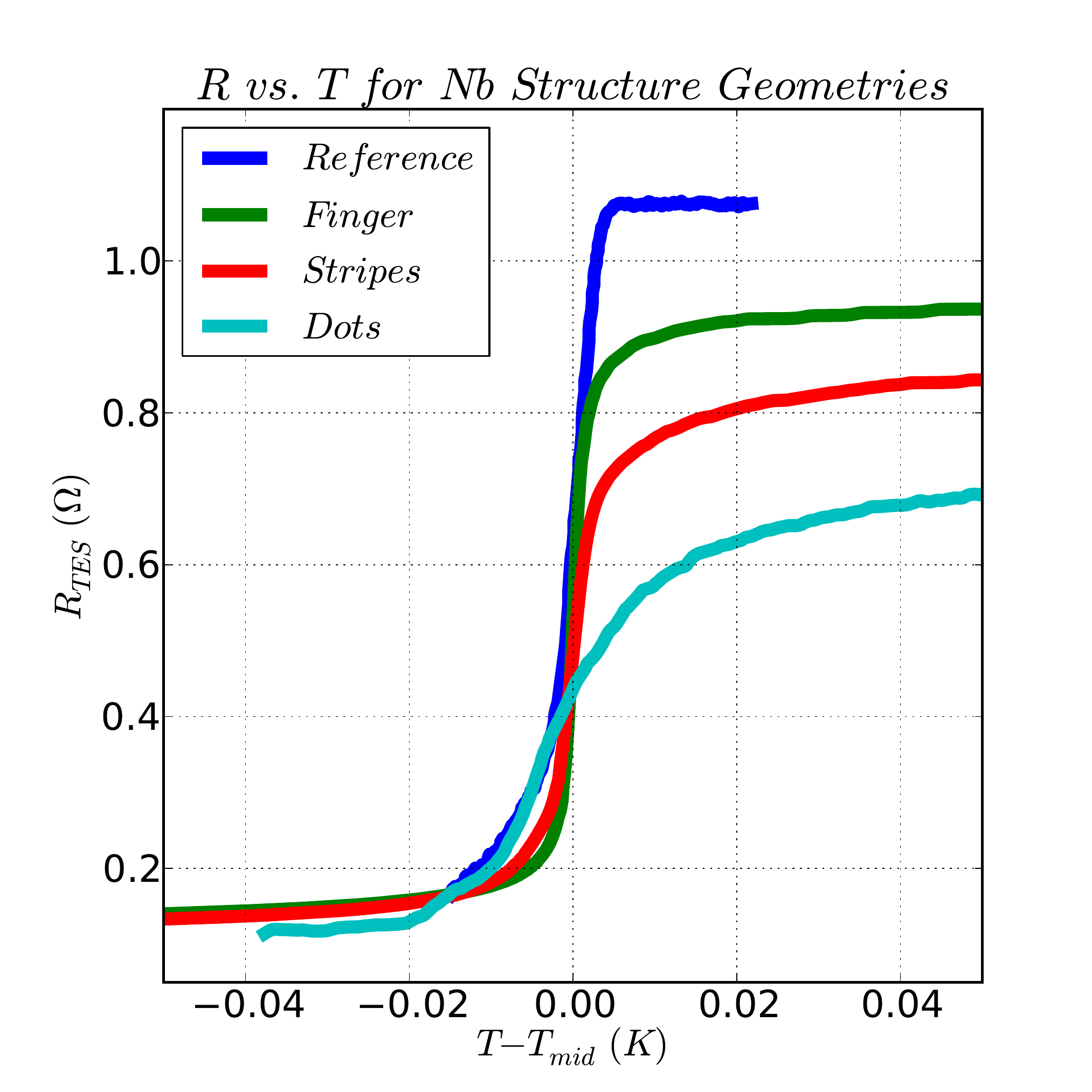}
   \end{tabular}
   \end{center}
   \caption[Effect of Nb Structures on R(T)] 
   { \label{fig:Nbstruct} 
Effect of Nb structures on the R(T) profile of prototype TESs.  The ``reference'' sample is a plain MoAu bilayer TES.  The ``finger'' pattern was a bar 11 $\mu$m wide that completely bisected the TES across its middle, perpendicular to the direction of current flow.  The ``zebra stripes'' were a series of 11 $\mu$m wide parallel bars partially spanning the TES.  Our final design was the ``dots'' arrangement of 3 $\mu$m diameter dots spaced 11 $\mu$m apart.}
   \end{figure}

Adding bling serves to increase $\tau_0$, however the bare MoAu TES has such a high $\alpha$ that the loopgain was too large for stable operation.  Our solution was to pattern Nb structures onto the TES itself, which leads to an interaction between the superconducting Nb and Mo/Au TES bilayer that widens the superconducting transition and reduces $\alpha$.  Figure \ref{fig:Nbstruct} shows R(T) profiles for a standard MoAu TES and test detectors with the addition of different Nb patterns on the TES.  Our final design consists of rows of 3 $\mu$m diameter dots with 11 $\mu$m separation, which provides stable performance with acceptable loopgains.

   \begin{figure}
   \begin{center}
   \begin{tabular}{c}
   \includegraphics[height=7cm]{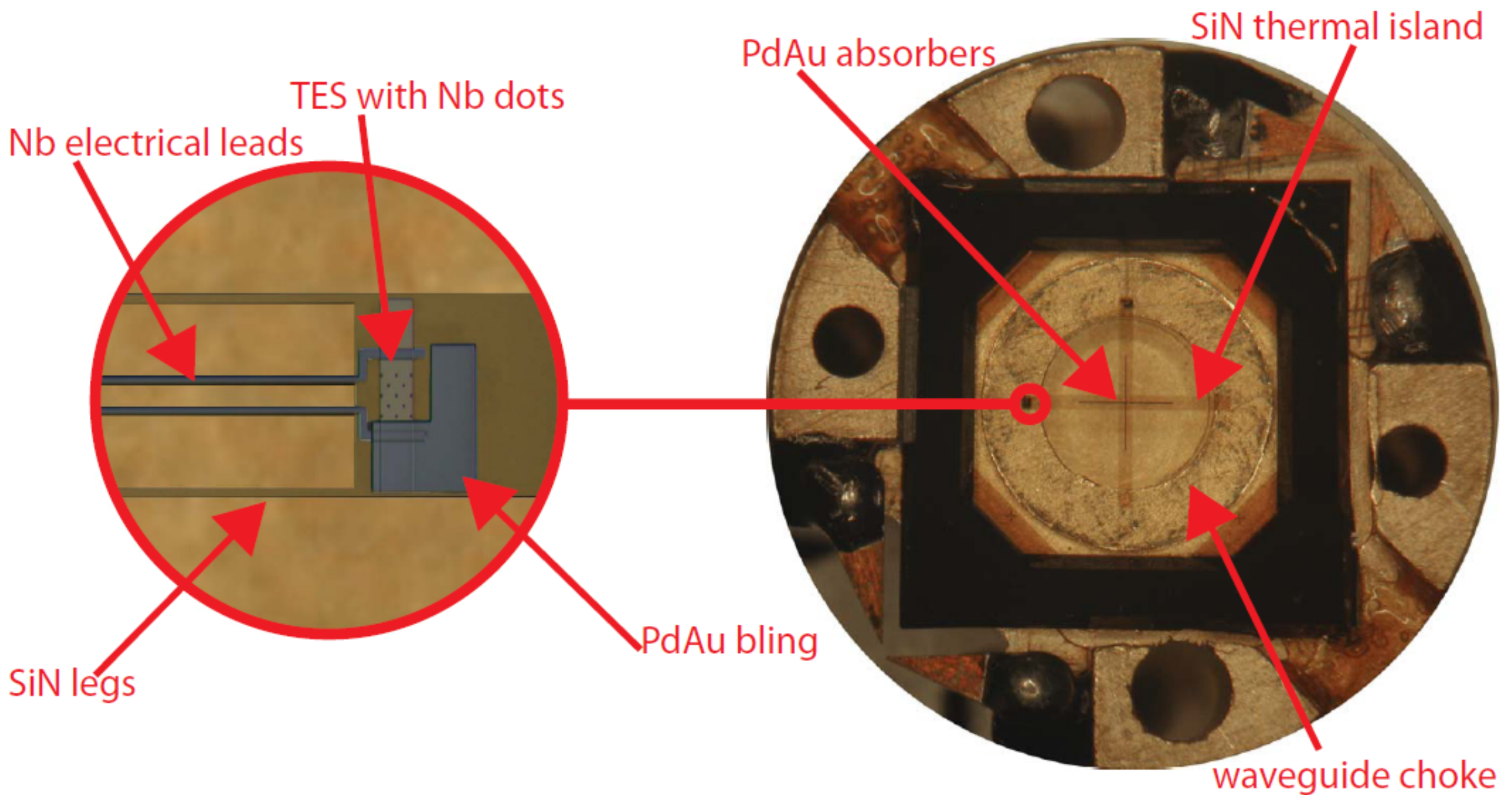}
   \end{tabular}
   \end{center}
   \caption[Pixel base] 
   { \label{fig:pix_base} 
A pair of single-polarization detctors mounted face-to-face and aligned $90^{\circ}$ to each other on a detector module base.  Each detector has a single TES, along with an L-shaped insert (called bling) to increase its heat capacity, deposited on the edge of the SiN thermal island.  Note the annular waveguide choke below the absorber membranes, into which the TES extends to minimize direct coupling to the waveguide fields.  Distance between the two detectors is maintained by laying 25 $\mu$m-diameter Al wirebonds along the Si frames.}
   \end{figure}

\section{CHARACTERIZATION OF 90 GHz PIXELS} 
\label{sec:Characterization}
\subsection{Optical Measurements}

\begin{figure}
   \begin{center}
   \begin{tabular}{c}
   \includegraphics[height=7cm]{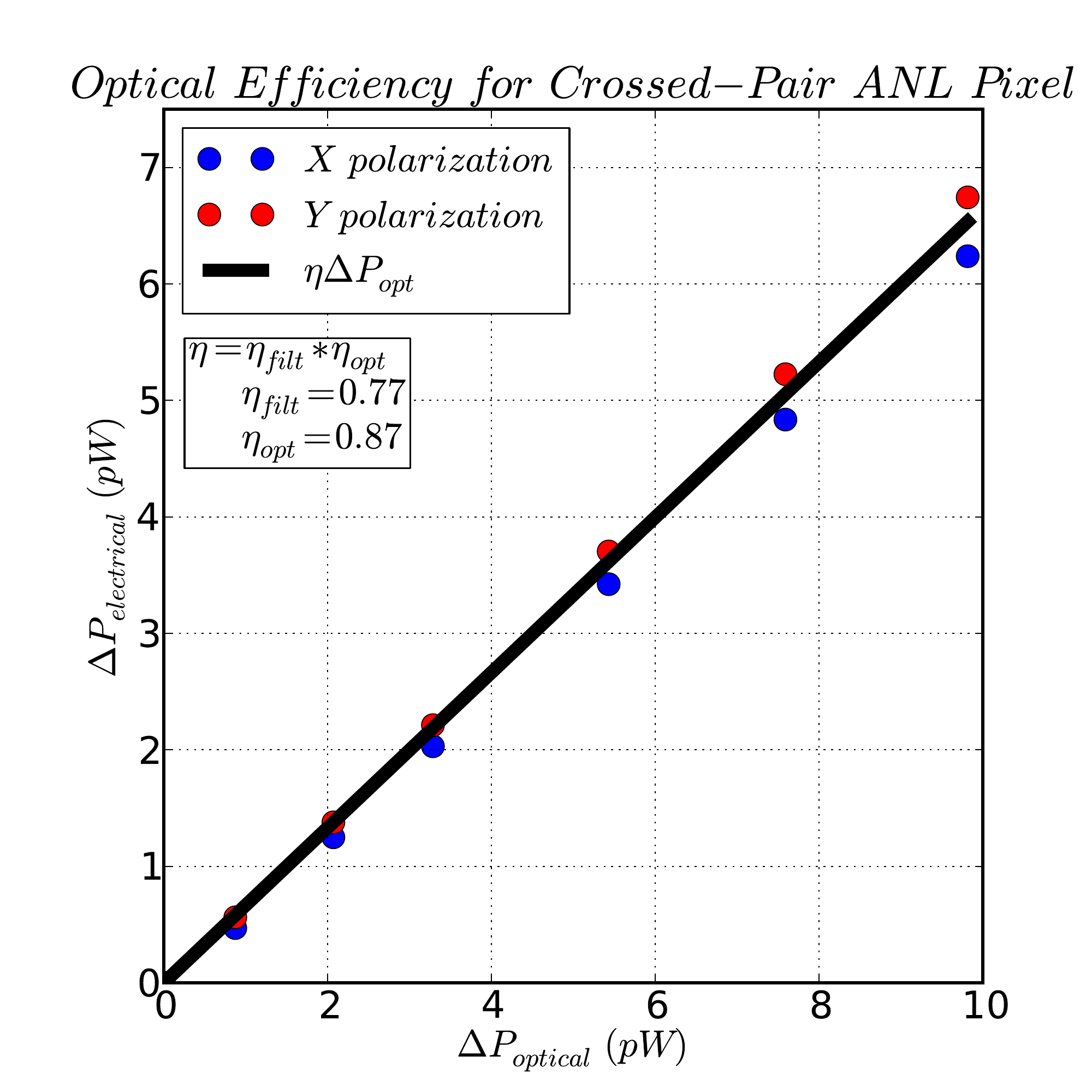}
   \end{tabular}
   \end{center}
   \caption[Polarization Efficiency] 
   { \label{fig:OptEff} 
Optical coupling efficiency of a 90 GHz pixel to a  blackbody load with a highly emissive coating, stepped in temperature from 4 K to 30 K.  The temperature of the optical load is changed and the electrical power required to reach a constant point in the TES transition is measured.  Since the total power to reach a given $T_{TES}$ is fixed, the change in electrical power is the opposite of the optical coupling efficiency multiplied by the emitted optical power from the cold load.  Here the values of $\Delta P$ are the change in power relative to the 4 K base temperature of the cold load, when optical power is minimized and electrical power is at a maximum.}
   \end{figure}

\begin{figure}
   \begin{center}
   \begin{tabular}{c}
   \includegraphics[height=7cm]{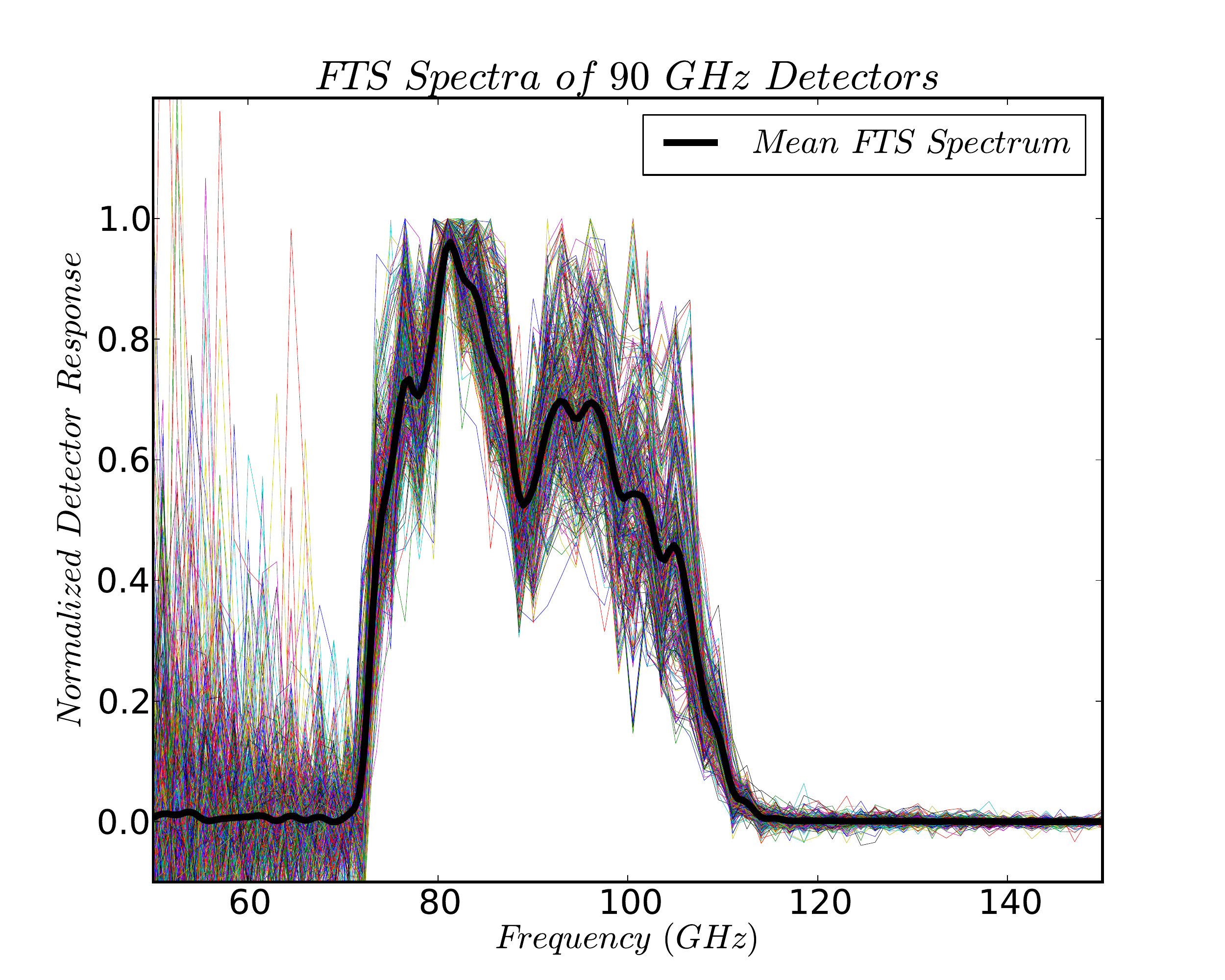}
   \end{tabular}
   \end{center}
   \caption[FTS Band] 
   { \label{fig:Band} 
Individual and array-averaged response spectra of the SPTpol 90 GHz detectors.  Data was taken with a Fourier transform spectrometer at the South Pole.  The band is a reasonable match to the target design.}
   \end{figure}

Optical coupling efficiency was measured by modulating the temperature of a beam-filling blackbody load and measuring the corresponding electrical power in the TES transition.  At a constant stage temperature, the total dissipated power at a given point in the transition is constant, and is composed of $P_{ELEC}+\eta P_{OPT}$, where $\eta$ is the total coupling efficiency to the emitted optical power from the load.  We account for a transmission efficiency factor $\eta_{filt}$  resulting from the imperfect transmission through the various band-defining and heat blocking filters, as well as the optical coupling efficiency factor $\eta_{opt}$, which parameterizes the coupling of the detectors to the optical power incident on the feedhorn aperture.  Assuming unity emissivity of the blackbody load and 99$\%$ transmission of one infrared blocking filter, 95$\%$ transmission of five band-defining and harmonic blocking filters, the total filter transmission is 77$\%$.  As shown in Figure \ref{fig:OptEff}, the optical coupling efficiency assuming single-moded and diffraction limited performance of the feedhorn and waveguide, is then 87$\%$, which is within expectations based on HFSS simulations of the band-averaged absorber-to-waveguide coupling \cite{McMahon}.  The transmission band from a Fourier transform spectrometer (FTS) illuminating the deployed focal plane is shown in Figure \ref{fig:Band}.

\begin{figure}
   \begin{center}
   \begin{tabular}{c}
   \includegraphics[height=7cm]{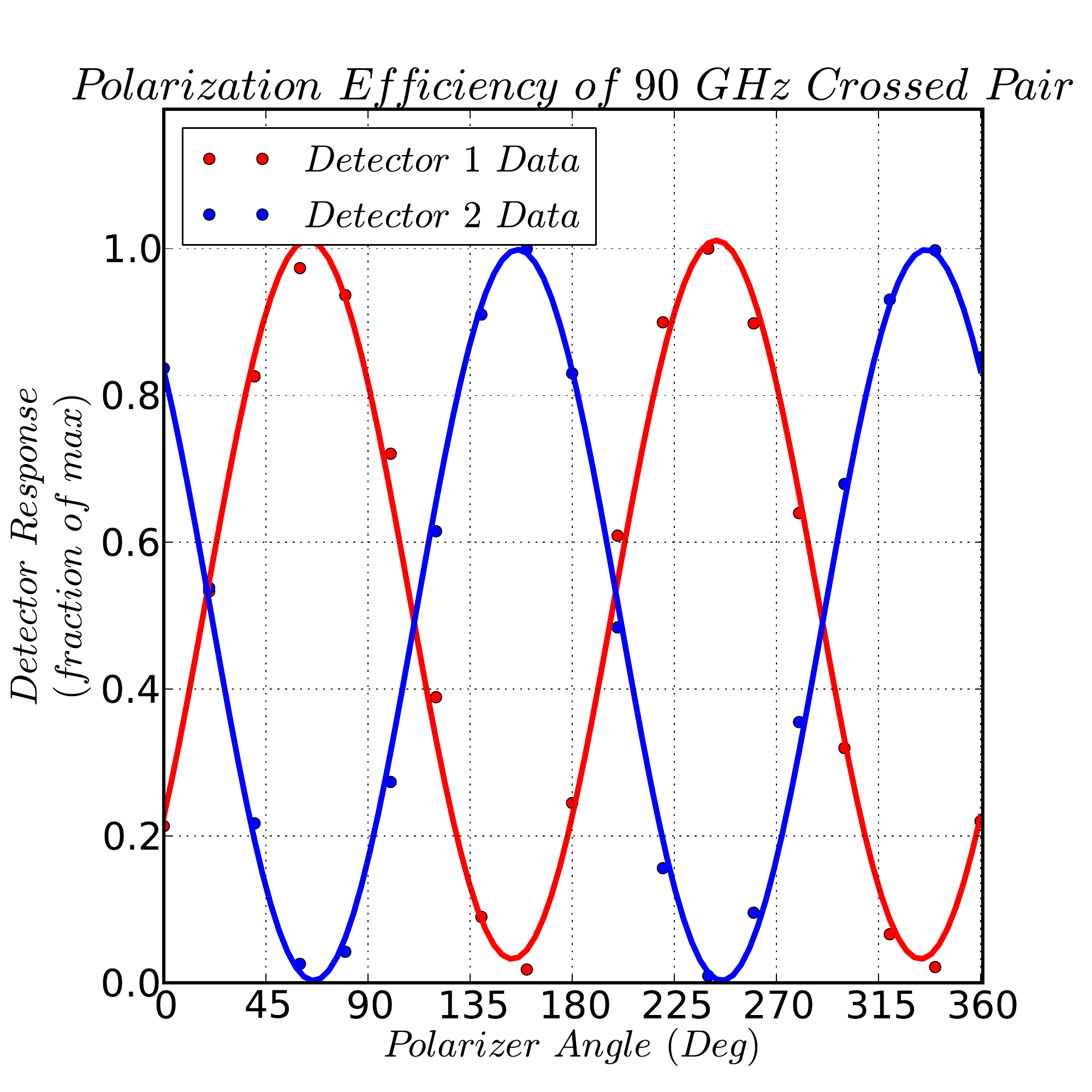}
   \end{tabular}
   \end{center}
   \caption[Polarization Efficiency] 
   { \label{fig:PolEff} 
Polarization response as a function of rotating polarizer grid angle.  Two crossed bolometers are shown.  The polarizer was rotated through $360^{\circ}$ in front of a chopped signal and the response of both detectors was recorded.  Each plot is normalized to the maximum response of that detector.  In both detectors, the maximum measured response to a crosspolar signal was $<1.6 \%$ of the full copolar response. }
   \end{figure}

Optical loading while looking into a 300 K load, as in laboratory testing, drives these detectors above their transition.  However, our final 90 GHz TES design with Nb dots and PdAu bling included a small but significant R(T) slope even in the normal regime, so some measurements, including polarization efficiency, were made using the normal-detector response to a large signal.  We illuminated the detectors with a chopped signal sent through a polarizing grid assumed to be 100$\%$ efficient.  The response to a strong chopped signal as a function of grid angle was recorded for two crossed detectors.  In this case,  one detector was successfully biased in its transition due to an anomolously high $\kappa$ resulting from accidental glue deposition on one of the thermal legs, while the other was in its normal regime but with responsivity sufficient for a high signal-to-noise measurement.  The results as shown in Figure \ref{fig:PolEff} are normalized to the maximum response from each detector and plotted with a linear fit to $A cos(\Theta_{p})+B sin(\Theta_{p}) + C$, where $\Theta_p$ is the polarizer angle.  From the fits, both detectors show a cross-polarization response of $< \,1.6 \%$ of the maximum copolar response, and a relative alignment of $90.6^{\circ}$ \cite{Chang}.

\subsection{TES Electrothermal Properties}

\begin{figure}
   \begin{center}
   \begin{tabular}{c}
   \includegraphics[height=7cm]{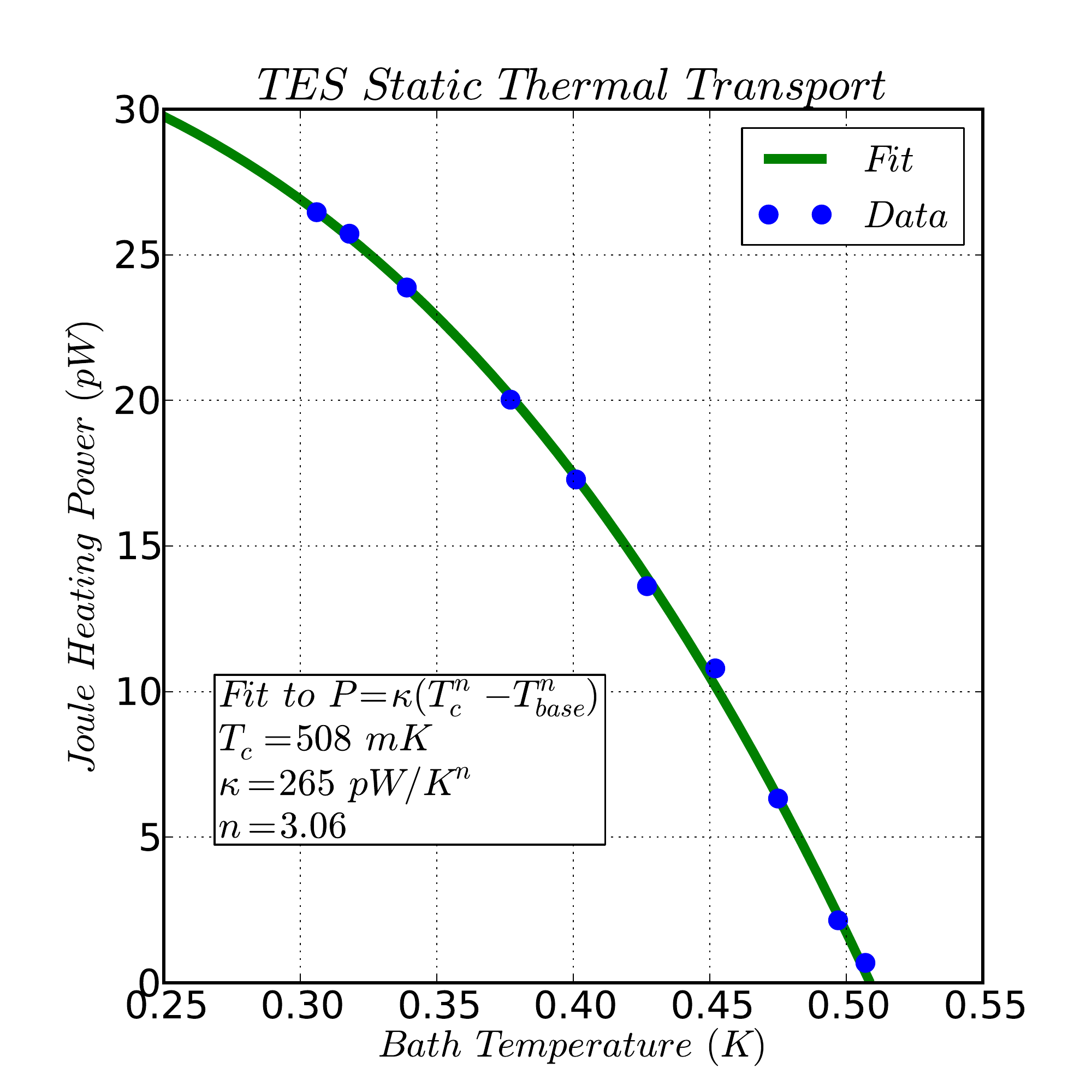}
   \end{tabular}
   \end{center}
   \caption[Fit to G and $T_C$] 
   { \label{fig:Gfit} 
Fit of $P=\kappa (T_C^n-T_{bath}^n)$ to transition powers at varied base temperature.  Transition temperature ($T_C$) from the fit to the model is near the 510 mK target, saturation power at 300 mK bath temperature is ~25 pW, and the form of heat transfer (parameterized by n) is consistent with expectations for the SiN legs. }
   \end{figure}

The static TES thermal parameters $T_C$ and $G_{dyn}$ can be measured together by using a series of TES current vs. bias voltage (I-V) load curves taken at various bath temperatures with a negligible optical load.  Optical loading is insignificant if the detectors are inside a light-tight 300 mK enclosure, and see less than 10$\%$  of the total design loading if their beams include only the inside of the 4 K thermal enclosure.  Each I-V curve sees the detector pass through its superconducting transition at a roughly constant electrical power, a result of a decrease in both input bias voltage and TES resistance.  Using the electrical power required to drive the detector to $T_C$ as a function of bath temperature, we fit for the static heat transfer equation (\ref{eqn:heattrans}).  Figure \ref{fig:Gfit} shows the $P(T_{base})$ data and the resultant fit.  We find a good agreement with the target thermal parameters of 510 mK $T_C$, 25 pW $P_{sat}$, and thermal conductivity exponent n=3, consistent with phonon transport through the SiN legs.

\begin{figure}
   \begin{center}
   \begin{tabular}{c}
   \includegraphics[height=7cm]{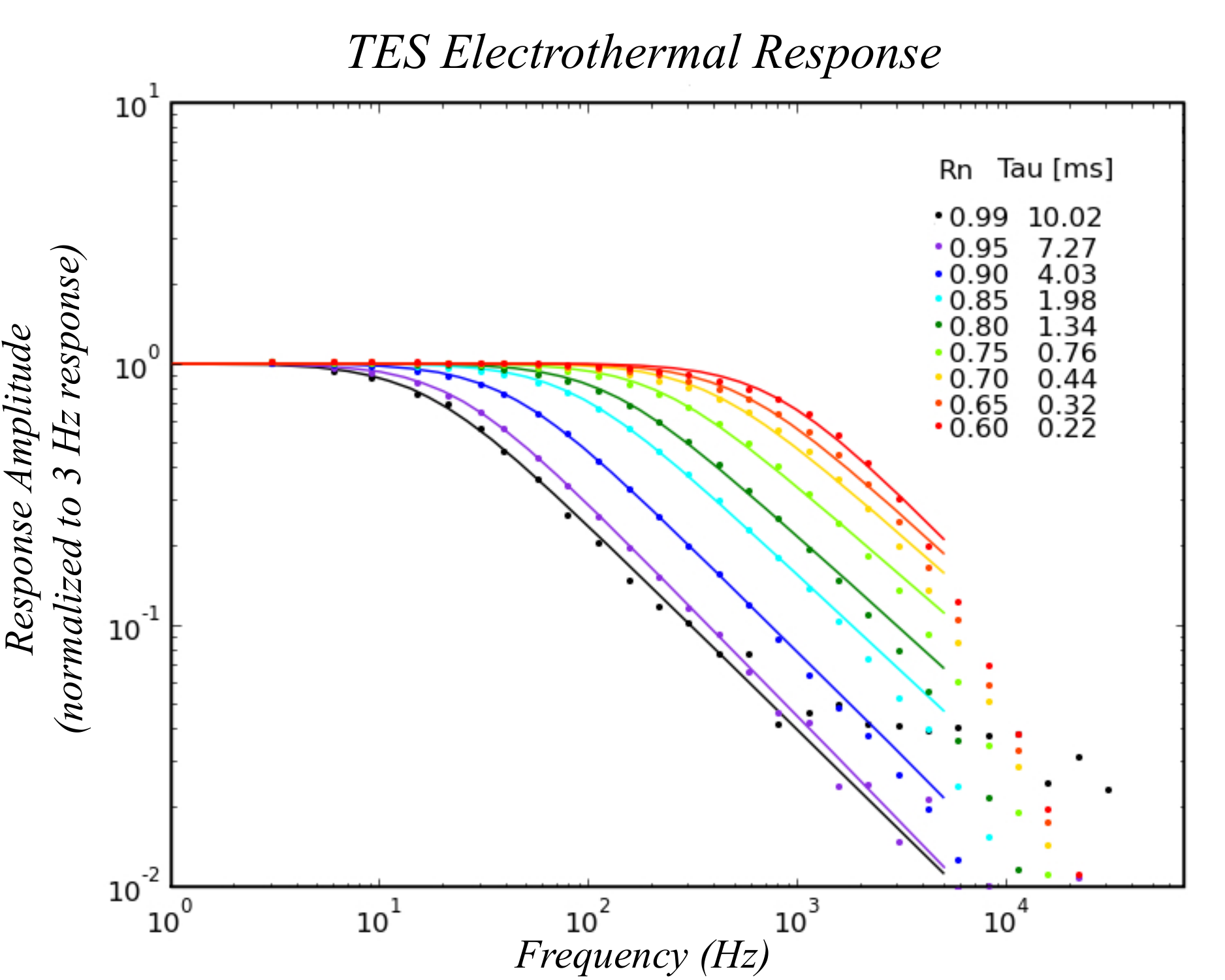}
   \end{tabular}
   \end{center}
   \caption[Electro-thermal response] 
   { \label{fig:ETF} 
Electro-thermal response as a function of frequency of thermal modulation and depth in the transition. Rn is the fractional depth in the transition, relative to $ R_{normal}$, while ``Tau[ms]'' is the time constant interpolated from the data.}
   \end{figure}

Electro-thermal stability of the TES is a key element in the performance of production detectors.  The ideal measurement of ETF is the TES response to incident optical power fluctuations at different frequencies.  However, the low-G design of the SPTpol 90 GHz detectors precludes such a direct measurement.  Our solution is to use a technique called the ``Lueker tickle''\cite{Lueker} to mimic a modulated optical load through use of the eletrical bias system.  The tickle utilizes a second bias channel with a small amplitude, $\delta P$, and a frequency offset from the carrier frequency by $\delta F$.  When the two signals are combined and sent to the TES, the resulting eletrical bias power beats at frequency $\delta$F and with an input amplitude $\delta P$.  We sweep $\delta F$ over a range from 3 Hz - 10 kHz and record the TES response amplitude as a function of the input beat frequency, normalizing to the detector response at 3 Hz.  Figure \ref{fig:ETF} shows an example of this measurement, illustrating the change in electro-thermal response as a function of depth in the transition (i.e. at various values of $\alpha$), and tickle frequency.  We can use the time constant of single-pole low pass filter fit to the data as a measurement of the loopgain of our detectors, according to Equation (\ref{eqn:LTau}).

\begin{figure}
   \begin{center}
   \begin{tabular}{c}
   \includegraphics[height=7cm]{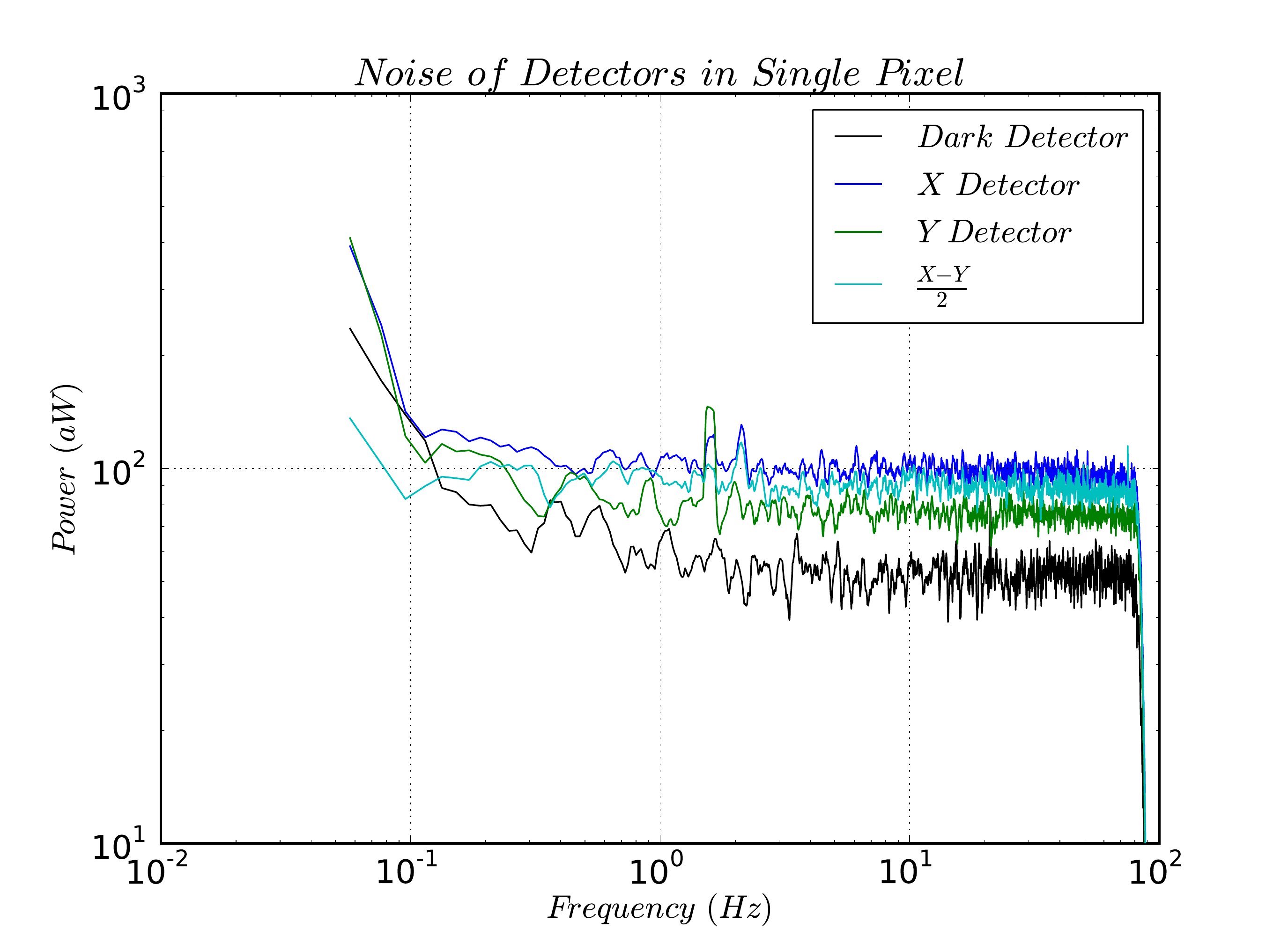}
   \end{tabular}
   \end{center}
   \caption[On-sky noise] 
   { \label{fig:noise} 
Noise spectra of X and Y detectors in a deployed pixel and the their differenced timestream, along with a dark spectrum from a different pixel taken during pre-deployment testing.  On-sky data was taken with the telescope stationary at an elevation within our observational field, while dark data was taken with the detector open to a light-tight 300 mK enclosure.}
   \end{figure}

Noise performance is critical to achieving the sensitivity required to extract a meaningful signal from CMB polarization.  The basic noise model for our detectors is a white noise component with a 1/f component at low frequency due primarly to slow modulations in atmospheric loading.  Each pixel's net polarization signal is derived from the difference of the individual bolometer timestreams, which ideally removes any non-polarized signal common to both detectors.  The white noise level is a combination of Johnson noise in the electrical readout circuit, thermal flucuation noise from the link between the TES and the bath, photon noise from the optical loading on the detector, and a small component of SQUID readout noise.  We seek to operate in the background-limited regime, where the overall noise is dominated by the photon noise component, which is approximately 70 $aW/\sqrt{Hz}$ for typical loading conditions.  In the background limit with current optical loading, our on-sky sensitivity is limited by the statistical fluctuations of incoming photons and thus can only be improved with more detectors and longer integration times.  Figure \ref{fig:noise} shows a sample noise spectrum from a single pixel, with both X and Y polarized detectors and their difference, along with a reference dark spectrum taken from a different detector prior to deployment.  The dark spectrum consists of Johnson noise, thermal fluctuation noise, and a small readout noise component, while the on-sky spectra introduce the background loading noise.

\section{CONCLUSION}  
\label{sec:conclusions}
In January of 2012, we deployed a 768 pixel polarimeter on the South Pole Telescope as part of an upgrade to the SPTpol phase of operations.  The 90 GHz band will be observed with 180 crossed-detector modules comprised of a profiled feed and two independent polarization-sensitive detectors mounted face-to-face in a circular waveguide.  Each detector couples to CMB radiation via a resistive PdAu absorbring bar, which couples optical power to a SiN thermal island on which is mounted a Mo/Au bilayer transition edge sensor.  Temperature fluctuations in the CMB are converted into modulations in the TES resistance which are read out by SPTpol digital-frequency multiplexed electronics.  We find detector optical coupling efficiency of 87$\%$, consistent with similuations, along with excellent polarization efficiency to minimize leakage of temperature into polarization signals.  We also confirm background-limited noise performance with stable electro-thermal loopgains and TES $\alpha$ under nominal loading and bias conditions.

\appendix    

\acknowledgments    
Work at Case Western Reserve University, the University of California - Berkeley, the University of Colorado - Boulder, and the University of Chicago is supported by grants from the NSF (awards ANT-0638937 and PHY-0114422), the Kavli Foundation, and the Gordon and Betty Moore Foundation.  Work at Argonne National Lab is supported by UChicago Argonne, LLC, Operator of Argonne National Laboratory ("Argonne"). Argonne, a U.S. Department of Energy Office of Science Laboratory, is operated under Contract No. DE-AC02-06CH11357.  We would aslo like to acknowlege support from the Argonne Center for Nanoscale Materials.  The McGill authors acknowledge funding from the Natural Sciences and Engineering Research Council, Canadian Institute for Advanced Research, and Canada Research Chairs program. Matt Dobbs acknowledges support from an Alfred P. Sloan Research Fellowship.

% References

\bibliography{jtlist}   %>>>> bibliography data in JTlist.bib
\bibliographystyle{spiebib}   %>>>> makes bibtex use spiebib.bst

\end{document}